\documentstyle[11pt,aaspp4]{article}

\newcommand{\ea}{{\it et al.}}

\newcommand{\kms}{km~$\rm{s}^{-1}$}

\newcommand{\cc}{$\rm{cm}^{-3}$}

\newcommand{\sfrac}[2]{\,{}^{#1}\!/_{#2}}
\newcommand{\beq}{\begin{equation}}
\newcommand{\eeq}{\end{equation}}
\newcommand{\bdm}{\begin{displaymath}}
\newcommand{\edm}{\end{displaymath}}

\slugcomment{Submitted to the Astrophysical Journal }

\begin{document}

\title{The Magnetic Geometry of Pulsed Astrophysical Jets}

\author{T. A. Gardiner, A. Frank}
\affil{Dept. of Physics and Astronomy,\\
       University of Rochester, Rochester, NY 14627-0171}

\begin{abstract} 

{\bf Hypersonic, highly collimated, mass outflows ({\it jets}) are
a ubiquitous phenomena in astrophysics.  While the character of the 
jets differ, many exhibit some form of quasi-periodic clumping indicating the
jet source is episodic or {\it pulsed}. The presence of pulsed jets in
so many astrophysical contexts suggests a common formation mechanism.
Such a process seems to have been found in {\it Magneto-centrifugal
launching}, the combination of magnetic and centrifugal forces that
occurs when a magnetized gaseous accretion disk orbits a central
gravitating source.  Observations of strong magnetic fields in jets are,
however, rare or indirect. Thus the presence and effects
of magnetic fields in YSO jets remains an unresolved issue of the
highest importance. In this letter we focus on what should be expected
of the structure of the fields in pulsed YSO jets.  We show that
combining velocity variability with an initial field configuration
consistent with collimated, Magneto-centrifugally launched jets leads
to a clear set of predictions concerning the geometry and relative
strength of the magnetic field components in evolving YSO (and perhaps
other) jets.}

\end{abstract}

\section{INTRODUCTION}

Hypersonic, highly collimated, mass outflows ({\it jets}) are
a ubiquitous phenomena in astrophysics. They occur in environments
as diverse as newly forming stars, (\cite{Reipurth97}) highly evolved
stars, (\cite{SokLiv94}) neutron stars and stellar mass black holes,
(\cite{Spruit00}) and massive black holes associated with the centers
of galaxies (\cite{Leahy91}).

Accretion disks are believed to play a key role in the physics of jet
launching.  Infalling, rotating matter is stored in these disks
until dissipation allows material to spiral inward and feed the
central, gravitating object.  Both YSO and AGN disks are believed
to support strong, well ordered magnetic fields.  The current
consensus holds that the these fields are the agents for producing
jets in a process known as {\it Magneto-centrifugal} launching.
In this mechanism, plasma in the disk is loaded on to co-rotating
field lines.  If conditions such as the inclination angle of the
field are favorable and a sufficient amount of energy is available,
the plasma is centrifugally flung outward along the field lines.
As the field is dragged backwards by the inertia of the plasma, strong
toroidal field components are generated which collimate the outflow
into a narrow jet.  We note however that the external medium might
also help focus the outflow. This process has been studied in detail
by many authors both analytically (\cite{HevNor89}, \cite{Pudritz91},
\cite{Shuea94}, \cite{Leryea99}) and through numerical simulations
(\cite{OP97a}, \cite{Romanova98}, \cite{Kudohea98})

Our goal in this letter is to show the effect of periodic velocity 
pulsations on the field geometry in the jet. In our studies of the effect of 
magnetic fields on jet propagation
(\cite{Frankea98,Frankea00,Gardinerea00}) we have found a simple,
universal relationship between velocity variability at the jet source
(the ``nozzle'') and the evolution of the jet's magnetic field when the
initial field is helical as would be expected from Magneto-centrifugal
launching models. In section 2 we present analytical calculations
showing the form of the evolved geometry.  In section 3 we present numerical
simulations which confirm the analytical models.

\section{Analytical Model}

We begin assuming that the jet can be modeled using the equations
of ideal magnetohydrodynamics in cylindrical symmetry. Consider a
system of cylindrical coordinates $(r,\phi,z)$ with the $z$-axis
parallel to the outflow and an accretion disk in the $z=0$ plane.
In magneto-centrifugal launching models for the production of
astrophysical jets, a magnetic field is anchored to a differentially
rotating accretion disk.  Plasma loaded on to the field lines is
``flung'' off the disk forming a wind (\cite{PudKon00}). Differential
rotation and plasma inertia then act to enhance the development of
a toroidal field component ($B_\phi$) whose pinch forces eventually
collimate the wind into a narrow jet.  Thus the general geometry
for the magnetic field in a steady, cylindrically symmetric,
magneto-centrifugally launched and collimated jet is that of a helix,
${\bf B} = (0,B_\phi(r), B_z(r))$( \cite{HeyNor89,Leryea98}).

The field is frozen-in to the beam as it propagates.  To clarify how 
pulsing will modify such
helical field configuration, consider first the induction equation.  In
steady flow the poloidal components of the velocity and magnetic field
are parallel.  Thus for collimated flows, the column of plasma which
makes up the jet will have no radial components of velocity or
magnetic field ($B_r = v_r = 0$). Assuming this remains approximately
unchanged during pulse propagation, the induction equation takes the
form,
\beq
\frac{\partial B_\phi}{\partial t} +
\frac{\partial}{\partial z} (v_z B_\phi - v_\phi B_z) = 0.
\label{induc}
\eeq
While $B_\phi$ and $B_z$ may be comparable (depending on initial
conditions in the disk and the exact nature of the launching
mechanism), the rotation rate in a jet will always be much smaller
than propagation speed ($v_z \gg v_\phi$).  Equation \ref{induc} is
thereby approximated by a simple conservation law with the same form
as the continuity equation,
\beq
\frac{\partial \rho}{\partial t} +
\frac{\partial}{\partial z} (\rho v_z) = 0.
\eeq
Thus, in a reference frame moving with the plasma, the ratio of the
toroidal field strength $B_\phi$ to the matter density $\rho$ is
approximately independent of time.

We now consider the effect of pulsing on the jet beam dynamics.
Assuming the jet beam remains approximately in equilibrium in the
radial direction it is possible to develop a simple 1-D kinematic
model for a jet with variable outflow speed 
(\cite{Gardinerea00,Smith1997,Raga1992}).
The essential details of this analysis are captured in Burgers'
equation.  When a flow exiting the nozzle of a jet varies with time,
the variations propagate downstream and are modified by compression
and rarefaction effects.  Consider material ejected at the jet nozzle
at some time $t_\circ$ with velocity $v_j(t_{\circ})$.  If
$v_j'(t_{\circ}) > 0$ (where the prime denotes a time derivative) then
material ejected at $t_\circ$ is moving faster than material ejected
earlier. The fast gas parcels encroach upon slower moving material
creating regions of compression. If $v_j'(t_{\circ}) < 0$ then
material ejected at $t_\circ$ is moving slower than material ahead of
it.  Gas parcels will move apart creating rarefaction regions of low
density.  If the velocity variations are strong enough compressive
regions will eventually steepen to form pairs of shock waves (each
pair containing an upstream and downstream facing shock). When the
variations are periodic a succession of shock pairs will propagate
down the beam with each pair separated by a rarefaction region.

In regions of smooth flow, away from shocks, we may combine the
continuity equation with Burgers' equation to obtain a relation for
the density of a gas parcel $\rho$ at time $t$ which was launched at
time $t_{\circ}$ with velocity $v_j(t_{\circ})$.  Combining this
result with the proportionality of the density and toroidal magnetic
field component we express the evolution of the field geometry as
\beq
\frac{B_\phi}{B_z} = \left( \frac{B_{\phi,o}}{B_z} \right) 
\left( \frac{1}{1-\kappa(t-t_{\circ})} \right)~~\textrm{where}~~
\kappa = \left(\frac{v_j'(t_{\circ})}{v_j(t_{\circ})}\right).
\label{fieldcomp}
\eeq
In the equation above we have used the fact that the poloidal
component $B_z$ lies parallel to the gas motions and is independent of
time.  Equation \ref{fieldcomp} describes how the helical magnetic
field lines are compressed or stretched respecitively by increasing or
decreasing velocity variations at the nozzle.  While equation
\ref{fieldcomp} breaks down when compressive regions steepen into
shocks, consideration of the MHD jump conditions allows the post-shock
field conditions to be calculated.  When post-shock cooling is
included the compression of the toroidal field can become quite large
$B_\phi/B_{\phi,o} > 10$.  Thus from simple dynamical arguments we
infer that a pulsing jet with an embedded helical field will
inevitably develop a periodic field structure consisting of rarefied
regions where the helix is {\it combed out} and has a more poloidal
geometry alternating with denser, toroidally dominated {\it knots}.
The rarefaction regions may become poloidally dominated depending on
the initial value of $\frac{B_\phi}{B_z}$ and the age of the jet (the
ratio asymptotically decreases as $1/\kappa t$).

\section{Numerical Simulations}

To confirm the predictions of our simple 1-D model we have performed
axisymmetric (2.5-D) numerical simulations of a pulsed radiative MHD
jet.  The conditions in our model are appropriate to protostellar jets
with an initial number density, velocity, temperature and sonic Mach
number in the jet of 120 \cc, 200 \kms, $2.5\times10^3$ K and 34
respectively.  The jet was initialized with a dynamically weak helical
field (plasma $\beta\approx 10\rightarrow10^3$ and
$\frac{B_{\phi,o}}{B_{z,o}}\approx 0\rightarrow10$).  The toroidal
field strength varies with an approximately paraboloidal shape as a
function of radius peaking at $\sfrac{1}{2}$ the jet radius.  The
$z$-component of the magnetic field was chosen to be uniform and
permeating both the jet and ambient medium.  The pressure and tension
forces associated with the toroidal field in the jet are balanced by
the gas pressure.  The velocity at the jet nozzle was varied with a
sinusoidal function ($v_j(t) = v_o [1 + A\sin(\omega t)]$) where $A =
0.25$ and the pulsation period was 80 years.  The jet was driven into
a constant ambient medium with density and temperature 60 \cc and
$5\times10^3$ K respectively.  In Figures 1a and 1b we show the
density and magnetic field structure from the simulation.  The density
image shows the high density regions behind the bow shock at the head
of the jet.  In addition there are two high density knots in the beam
where compressive regions have steepened into shocks and a third
region of compression near the base of the jet.  The magnetic field
image shows that the intensity of the toroidal field is highest behind
the shocks.  In particular note the concentration of toroidal field in
the knots.  Note also that the $B_z$ component is relatively
unaffected by the dynamics except at the jet edges or where radial gas
motions are important, such as near the head of the jet.  In Fig 2 we
show a plot of $B_\phi(z)$ at two times in the evolution of the
simulation which differ by $\sfrac{1}{2}$ of a pulsation period.  The
growth of the toroidal field component as velocity pulses steepen into
shocks is clearly evident as is its rarefaction in the region between
the pulses. Note that our simulation shows a maximum increase of
$\frac{B_{\phi}}{B_{\phi,o}}\approx 30$.  From the plot it appears as
if the ratio of the maximum to the minimum toroidal field strength has
stabilized at $\approx 30$. Plotting a similarly scaled density
profile with the toroidal field strength shows the approximations made
to be quite acceptable.  In the rarefaction regions $B_\phi$ continues
to decrease at the end of the simulation.  Thus in proto-stellar jets,
whose lifetimes are two orders of magnitude larger we expect the
rarefaction regions to become poloidally dominated.

Thus the numerical simulations confirm the results of our simple model.
In flows with strong cooling, helical fields embedded in pulsed jets
will invariably lead to toroidally dominated knots separated by low
density regions which can be poloidally dominated.  Our model allows
a clean prediction of the field structure in velocity variable jets.
Polarization maps (the most promising means of determining field structure
, \cite{Rayea,Chrysostomou,Minchinea95}) should show polarization 
vectors whose orientation angle changes by $90^\circ$ at the boundary
of bright knots.  Our model also allows one to cleave between
different models of the origin of knots based on the observed field
structure.  There is currently some debate over the origin of knots in
protostellar flows.  It remains unclear if the knots are due to
pulsing of the jet source or instabilities in the beam, though recent
observations make a strong case for pulsing in the case of jets from
young stars. \cite{Zinnecker} While shock waves will occur in a
hypersonic beam which becomes unstable they will be both weak and
oblique. \cite{Stone97} Thus instabilities are not likely to lead to
the strong toroidal/poloidal (compression/rarefaction) regions
predicted by our model.

\section{Conclusions}

We have shown that pulsating jets with initially helical fields
will evolve to an alternating poloidal/toroidal dominated field geometry.
The strongest toriodal field will be confined to the dense knots
in between the shock pairs while the rarefaction regions will contain
a dominant polodial field.

While much of the discussion in this paper has been concerned with the
geometry of the magnetic field, equally important is the magnetic
field topology.  It seems quite reasonable to expect that the
compresison of the toroidal magnetic field may be quite a violent and
unstable process on small scales.  To predict the resultant relaxed
field configuration one might attempt to apply Taylor's theory
\cite{Taylor86} for magnetic reconnection in plasmas.  While this is 
beyond the scope of this letter, possible consequences include
topological changes such as magnetic field reversals and reconnection.
These in turn should have observational as well as dynamical consequences.

Finally we note that while this work is most relevant to strongly
cooling flows such as those associated with YSOs (and Planetary
Nebulae) it may also be relevant to the more adiabatic flows
associated with AGN and Micro-quasars.  While the mass and field
compression will be limited to a factor of $4$ in an adiabatic flow
the rarefaction seen in our model will occur even without cooling.  
Thus, depending on initial conditions, the alternating
toroidal/poloidal geometry may occur in all pulsed jets.


%
\newpage
\begin{figure}
\figurenum{1a}
\caption{Logarithm of the density after the jet has propagated 
for 365 years.  The colorbar indicates the number density in units of 
Hydrogen atoms per cm$^3$.}
\end{figure}

\begin{figure}
\figurenum{1b}
\caption{Composite image showing the poloidal magnetic field lines on 
the top panel and the strength of the toroidal magnetic field component 
on the bottom panel after propagating for 365 years.  The colorbar indicates 
the toroidal field strength in Gauss.}
\end{figure}

\begin{figure}
\figurenum{2}
\caption{Logarithm of $B_{\phi}/B_{\phi,o}$ along 
the jet at $\sfrac{1}{2}$ of the jet radius.  The two evolutionary times 
chosen differ by $\sfrac{1}{2}$ of the pulsation period.}
\end{figure}


\newpage

\begin{thebibliography}{}
%
\bibitem[Chrysostomou \ea 2000]{Chrysostomou}
Chrysostomou, A., Gledhill, T. M., Ménard, F., Hough, J. H.,
Tamura, M., Bailey, J., 2000, MNRAS, 312, 103.
%
\bibitem[Frank \ea{} 1998]{Frankea98} 
Frank A., Lery T., Gardiner, T.,Jones T. W. \&  Ryu D. 1998, 
\apj, 494, L79.
%
\bibitem[Frank \ea{} 2000]{Frankea00}
Frank A., Ryu D., Jones T. W. \& Noriega-Crespo A. 1998, \apj, 494, L79.
%
\bibitem[Gardiner \ea{} 1999] {Gardinerea00}
Gardiner T., Frank A., Ryu D., Jones T., 2000, \apj, 530, 834.
%
\bibitem[Heyvaerts \& Norman 1989] {HeyNor89}
Heyvaerts, J \& Norman, C., 1989, ApJ, 347, 1055.
%
\bibitem[Heyvaerts \& Norman 1989]{HevNor89}
Heyvaerts J., Norman C.A.,  1989, \apj, 347, 1055
%
\bibitem[Kudoh \ea 1998]{Kudohea98}
Kudoh T., Matsumoto R., Shibata K.,  1998, \apj, 508, 186
%
\bibitem[Leahy 1991] {Leahy91}
Leahy J.P., 1991, in ``Beams and Jets in Astrophysics'', ed. P. Hughes, 
(Cambridge U. Press).
%
\bibitem[Lery \ea{} 1998] {Leryea98}
Lery T., Heyvaerts J., Appl S., Norman C.A., 1998, A\&A 337, 603.
%
\bibitem[Lery \ea 1999] {Leryea99}
Lery T., Heyvaerts J., Appl S., Norman C.A., 1999, A\&A, in press
%
\bibitem[Minchin ea 1995]{Minchinea95}
Minchin, N. R., Sandell, G., Murray, A. G. 1995, A\&A, 293, 61.
%
\bibitem[Ouyed \& Pudritz 1997]{OP97a} Ouyed R. \& Pudritz R. E. 1997a, 
\apj, 482, 712
%
\bibitem[Pudritz 1991]{Pudritz91}
Pudritz R.E. 1991, in ``The Physics of Star Formation and Early Stellar
Evolution'', eds. C.J. Lada and N.D. Kylafis, NATO ASI Series (Kluwer), 365
%
\bibitem[Pudritz \& Konigl 2000]{PudKon00}
Pudritz R., \& Konigl, A., 2000, in Protostars \& Planets IV.
%
\bibitem[Raga \ea{} 1992]{Raga1992} 
Raga, A. C. \& Kofman, L. 1992, \apj, 386, 222.
%
\bibitem[Romanova \ea 1998]{Romanova98} Romanova M. M., Ustyugova, G. V., 
Koldoba A.V., Chechetkin V.M., \& Lovelace R.V.E. 1998, \apj, 
500, 703
%
\bibitem[Ray \ea{} 1996]{Rayea}
Ray T.P., Mundt R., Dyson J.E., Falle S.A.E.G., Raga A.C., 1996, 
\apj, 468, L103.
%
\bibitem[Reipurth 1997]{Reipurth97}
Reipurth B., 1997, in Herbig-Haro Flows and the Birth of
Low Mass Stars, in IAU Symposium no. 182, eds B. Reipurth \& C Bertout
(Kluwer, Dordrecht)
%
\bibitem[Soker \& Livio 1994]{SokLiv94}
Soker N., \& Livio M., 1994, AJ 421, 219.
%
\bibitem[Spruit 2000]{Spruit00}
Spruit, H.C., 2000, in IAU Symposium 195 ed P. Martens, astro-ph/0003043.
%
\bibitem[Shu {\it et al.} 1994] {Shuea94}   
Shu F., Najita J., Ostriker E., Wilkin F., Ruden S., \& Lizano S.,
1994, \apj, 429, 781
%
\bibitem[Smith \ea{} 1997]{Smith1997} 
Smith, M. D., Suttner, G., \& Zinnecker, H. 1997, \aap, 320, 325.
%
\bibitem[Stone \ea{} 1997]{Stone97}
Stone, J., Xu, J., Hardee, P., 1997, \apj, 483, 136.
%
\bibitem[Taylor 1986]{Taylor86}
Taylor, J. B., 1986, Rev. Mod. Phys., 58, 741.
%
\bibitem[Zinnecker \ea{} 1998]{Zinnecker}
Zinnecker, H., McCaughrean, M. J., Rayner, J. T., 1998, Nature, 394, 862.
%
\end{thebibliography}
\end{document}